\documentclass[aip, jmp, showpacs, showkeys]{revtex4-1}
\usepackage{amsfonts}
\usepackage{amssymb}
\usepackage{amstext}
\usepackage{amsmath}
\usepackage{amsthm}
\usepackage{latexsym}
\usepackage{color}

\begin{document}

\makeatletter

\title{Symbolic methods for the evaluation of sum rules of Bessel functions}

\author{D. Babusci}
\email{danilo.babusci@lnf.infn.it}
\affiliation{INFN - Laboratori Nazionali di Frascati, via E. Fermi, 40, IT 00044 Frascati (Roma), Italy}

\author{G. Dattoli}
\email{dattoli@frascati.enea.it}
\affiliation{ENEA - Centro Ricerche Frascati, via E. Fermi, 45, IT 00044 Frascati (Roma), Italy}

\author{K.~G\'{o}rska}
\email{katarzyna.gorska@ifj.edu.pl}
\affiliation{H. Niewodnicza\'{n}ski Institute of Nuclear Physics, Polish Academy of Sciences, ul.Eljasza-Radzikowskiego 152, 
PL 31342 Krak\'{o}w, Poland}
\affiliation{Instituto de F\'{\i}sica, Universidade de S\~{a}o Paulo, \\
P.O. Box 66318, B 05315-970 S\~{a}o Paulo, SP, Brasil}

\author{K.~A.~Penson}
\email{penson@lptl.jussieu.fr}
\affiliation{Laboratoire de Physique Th\'eorique de la Mati\`{e}re Condens\'{e}e,\\
Universit\'e Pierre et Marie Curie, CNRS UMR 7600\\
Tour 13 - 5i\`{e}me \'et., B.C. 121, 4 pl. Jussieu, F 75252 Paris Cedex 05, France\vspace{2mm}}

\begin{abstract}
The use of the umbral formalism allows a significant simplification of the derivation of sum rules involving products of special functions  and polynomials. We rederive in this way known sum rules and addition theorems for Bessel functions. Furthermore, we obtain a set of new closed form sum rules involving various special polynomials and Bessel functions. The examples we consider are relevant for applications ranging from plasma physics to quantum optics.
\end{abstract}

%--------------------------------------
\maketitle
%-----------------------------------------------------------

\section{Introduction}
In this note we will show that the problem of deriving generating functions of special functions can be greatly simplified by the use of symbolic methods. This methods have been widely employed recently to deal with the Ramanujan master theorem \cite{GHHardy40, TAmdeberhan09, TAmdeberhan11} and to derive closed expressions for integrals involving various combinations of Bessel functions.

The methods we are going to describe in the paper complete previous researches summarized in \cite{DBabusci11, DBabusci12a}, where it was shown that the use of operational techniques allowed us the derivation of sum rules hardly achievable with conventional means.

We shall first present the main ingredients of the formalism. They will be employed in the following to establish known and new sum rules within a simplified and unifying point of view. We start by introducing the umbral notation with the operator $\hat{c}$:
\begin{equation}\label{eq1}
\hat{c}^{\alpha} \varphi_0 = \varphi_\alpha, \qquad \varphi_\alpha = \frac{1}{\Gamma(1 + \alpha)}, 
\end{equation}
$(\hat{c}^{\alpha + \beta} = \hat{c}^{\alpha}\, \hat{c}^{\beta})$, without any restriction on $\alpha$.

According to the previous definition, Bessel functions of the first kind \cite{GEAndrews01} can be written as
\begin{equation}\label{eq2}
J_{\alpha}(x) = \sum_{k = 0}^{\infty} \frac{(-1)^k\,(x/2)^{2k + \alpha}}{k!\, \Gamma(\alpha + k + 1)} = \left(\hat{c}\, \frac{x}{2}\right)^{\alpha}\, e^{-\hat{c}\, \left(x/2\right)^2}\, \varphi_0,
\end{equation}
while the Tricomi-Bessel functions of order $\alpha$ \cite{GEAndrews01, FGTricomi54, GDattoli04} are given by
\begin{equation}\label{BesTri}
C_{\alpha}(x) = \sum_{k = 0}^{\infty} \frac{(-x)^k}{k!\, \Gamma(\alpha + k + 1)} = x^{-\alpha/2}\, J_{\alpha}(2\sqrt{x}) = \hat{c}^{\alpha}\, e^{-\hat{c} x}\, \varphi_0. 
\end{equation}
For the two-variable Laguerre polynomials \cite{DBabusci10,DBabusci12}, the same notation yields,
\begin{equation}\label{eq3}
L_{n}(x, y) = n! \sum_{k = 0}^n \frac{(-x)^{k} y^{n-k}}{(n-k)!\, (k!)^2} = y^{n} L_{n}\left(\frac{x}{y}\right) = (y - \hat{c} x)^{n}\, \varphi_0, 
\end{equation}
where $L_{n}(z)$ is the ordinary Laguerre polynomial. Finally, for the Wright function \cite{RGoreflo99} we have
\begin{equation}\label{eq10}
W_{\nu}(x| \mu) = \sum_{r=0}^{\infty} \frac{x^{r}}{r!\, \Gamma(\nu + \mu\, r)} = \hat{c}^{\nu-1}\, e^{\hat{c}^{\,\mu} x} \varphi_0.
\end{equation}

Given a function with series expansion 
\begin{equation}\label{n1}
f(x) = \sum_{n=0}^{\infty} \frac{a_n}{n!}\,x^n,
\end{equation}
we will define the corresponding Laguerre-based function as \cite{DBabusci10}
\begin{equation}\label{eq4a}
_{L}f(x, y) = \sum_{n = 0}^{\infty} \frac{a_n}{n!}\,L_{n}(x, y).
\end{equation}
In an analogous way, we define the Hermite-based functions as ($m \in \mathbb{Z}_+$)
\begin{equation}\label{Her-based}
_{H}f(x, y) = \sum_{n = 0}^{\infty} \frac{a_n}{n!}\,H_{n}^{(m)}(x, y),
\end{equation}
with
\begin{equation}\label{eq4b}
H_n^{(m)}(x, y) = n! \sum_{k=0}^{[n/m]} \frac{x^{n-m k} y^k}{(n - m k)!\, k!}.
\end{equation}
For future reference, we remind that the two-variables Hermite polynomials verify the identity 
\begin{equation}\label{eq4c}
H_n^{(m)}(-x, y) = (-1)^n\,H_n^{(m)}\left(x, (-1)^m y\right)
\end{equation}
and their generating function is given by
\begin{equation}\label{eq4d}
\sum_{n=0}^{\infty} \frac{t^{n}}{n!}\, H^{(m)}_{n}(x, y) = e^{x t + y t^m}.  
\end{equation}
The results we will obtain in this paper will also be exploited to derive families of sum rules useful in classical and quantum optics. In particular, we will make a comparison with the conclusions of the recent 
analysis of \cite{GBevilacqua11}, in which the sum rules for the product of Bessel functions have been used to derive the expression for the work done by a modulated electric field forcing a damped harmonic oscillator.

\section{Generating functions of Bessel functions and special polynomials}

We prove the efficiency of the procedure by means of some introductory examples, which are a direct consequence of the previous definitions. We start with a well-known result (see Eq. (5.7.6.1) from 
\cite{APPrudnikov_2}). By using Eq. \eqref{eq2}, it is a simple task to show that
\begin{eqnarray}\label{Besumrul}
\sum_{n = 0}^{\infty} \frac{t^n}{n!}\,J_{n + \nu}(x) &=& \left(\hat{c}\,\frac{x}2\right)^\nu\,\exp\left[-\frac{\hat{c}}4\,(x^2 - 2\,x\,t)\right]\,\varphi_0 \nonumber \\
&=& \left(\frac{x}{x - 2 t}\right)^{\nu/2}\,J_{\nu}(\sqrt{x^2 - 2 x t}), \qquad \qquad (|2 t| < x, \quad \nu\in\mathbb{C}),
\end{eqnarray}
obtained here in a nonconventional and very straightforward way.

Furthermore, one has
\begin{equation}\label{n2}
\sum_{n=0}^{\infty} \frac{t^n}{n!}\,J_{m\,n}(x) = \,_{H}C_0^{(m)}\left(\frac{x^2}4, \left(-\frac{x}2\right)^m t \right),
\end{equation}
where 
\begin{equation}\label{Cfunc}
\,_{H}C_{\nu}^{(m)}(u, v) = \sum_{k = 0}^{\infty} \frac{(-1)^k}{k!\,\Gamma(\nu + k + 1)}\,H_k^{(m)}(u, v).
\end{equation}
are the so-called Hermite-based Tricomi functions \cite{DBabusci10}. To get the above sum rule, Eqs. \eqref{eq4c} and \eqref{eq4d} have been used, along with the following identity
($m \in \mathbb{Z}_+$)
\begin{equation}\label{eq7}
e^{- \hat{c}\,x + \hat{c}^m\,y} \varphi_0 = \left[\sum_{n = 0}^{\infty} \frac{\hat{c}^{\,n}}{n!}\,H_n^{(m)}(- x, y)\right]\,\varphi_{0} = \,_{H}C_0^{(m)} \left(x, (-1)^m\,y\right),
\end{equation}
that proves to be useful also to show that 
\begin{equation}\label{eq11}
\sum_{n = 0}^{\infty} \frac{t^n}{n!}\, J_{n/m}(x) =\,_{H}W_0^{(m)} \left( t \left(\frac{x}{2}\right)^{1/m}, -\frac{x^{2}}{4} \Big|\, \frac{1}{m} \right),
\end{equation}
where we have introduced the new functions
\begin{equation}\label{n3}
\,_{H}W_{\nu}^{(m)}(u, v |\mu) = \sum_{k = 0}^{\infty} \frac{H_k^{(m)}(u, v)}{k!\,\Gamma(\mu\,k + \nu + 1)},
\end{equation}
i.e., the Hermite-based functions obtained for $a_n = 1/\Gamma(\mu\,n + \nu + 1)$.

By using the result \eqref{eq3} we can easily derive the new (hardly achievable with conventional means) sum rule 
\begin{equation}
\sum_{n=0}^{\infty} \frac{t^n}{n!}\,J_n (z)\,L_n (x, y) = \,_{L}C_{0} \left(- \frac{x\,t\,z}2, \frac{z\,(z- 2\,y\,t)}4\right), 
\end{equation} 
where the Laguerre-based function $_{L}C_{\nu}$ is obtained from Eq. \eqref{Cfunc} by replacing $H_n^{(m)}$ with $L_n$. In the same way we can obtain a generating function involving the product of Hermite and Laguerre polynomials, which can be written as 
\begin{equation}
\sum_{n=0}^{\infty} \frac{t^n}{n!}\,L_n (x, y)\,H_n^{(2)} (z, w) = e^{y\,t\,(z + y\,w\,t)}\,_{H}C_0^{(2)}\left(x\,t\,(z + 2\,y\,w\,t), x^2\,w\,t^2\right), \label{eq9}
\end{equation}
through the use of Eq.~\eqref{eq7}.

\section{Symbolic derivation of some addition theorems}\label{s:symbolic}
Our method allows us to prove the Graf's addition theorem \cite{GEAndrews01} in a very straightforward way. By using Eq. \eqref{eq2} and the identity  
\cite{APPrudnikov_2}
\begin{equation}
\sum_{n = -\infty}^{\infty} t^n\,J_n (z) = \exp\left[\frac{z}2\,\left(t - \frac1{t}\right)\right], 
\end{equation}
we get the formula $(\nu \in\mathbb{C})$
\begin{equation}\label{eq14}
G^{(\nu)} (x, y; t) = \sum_{n = -\infty}^{\infty} t^n\,J_{n + \nu} (x)\,J_n (y) = \left(\hat{c}\,\frac{x}2\right)^\nu\,
e^{- \xi\,\hat{c} - \chi\,\hat{c}^{- 1}}\,\varphi_0
\end{equation}
where
\begin{equation}
\xi = \frac14\,(x^2 - x\,y\,t) \qquad\qquad \chi = \frac14\,\frac{y}{x\,t}. \nonumber
\end{equation}
After a simple manipulation, we can write 
\begin{equation}
G^{(\nu)} (x, y; t) =  \left(\frac{x}{2\,\sqrt{\xi}}\right)^\nu\,\sum_{k = 0}^{\infty} \frac{(- \sqrt{\xi}\,\chi)^k}{k!}\,J_{\nu - k} (2\,\sqrt{\xi}). 
\end{equation}
and, since (see Appendix)
\begin{equation}\label{Besumrul2}
\sum_{n = 0}^{\infty} \frac{(- t)^n}{n!}\,J_{\nu - n} (x) = \left(\frac{x - 2\,t}{x}\right)^{\nu/2}\,J_{\nu}(\sqrt{x^2 - 2\,x\,t}), \qquad \qquad (|2 t| < x,\quad \nu\in\mathbb{C}),
\end{equation}
one obtains
\begin{equation}\label{Gnu}
G^{(\nu)} (x, y; t) = \left(\frac{x - y\,t^{- 1}}{x - y\,t}\right)^{\nu/2}\,J_\nu \left(\sqrt{x^2 + y^2 - x\,y\,(t + t^{- 1})}\right), \quad (x > y/t).
\end{equation}
Then, setting $t = e^{i\,\theta}$, the Graf's addition theorem results:
\begin{equation}\label{eq18}
\sum_{n = -\infty}^{\infty} e^{i\,n\,\theta}\,J_{n + \nu} (x)\,J_n (y) = \left(\frac{x - y\,e^{- i\,\theta}}{x - y\,e^{i\,\theta}}\right)^{\nu/2}\,
J_\nu \left(\sqrt{x^2 + y^2 - 2\,x\,y\,\cos\theta}\right).
\end{equation}

In the same way, we can evaluate the following extension of the Neumann addition theorem
\begin{equation}\label{eq40}
\sum_{n = -\infty}^\infty t^n J_n (x)\,J_{2 n} (y) = \,_{H}K_0^{(2)}\left(\frac{y^2}4, \frac{x\,y^2\,t}8 \Big| \frac{2\,x}{y^2\,t}\right),
\end{equation}
where we have introduced the hybrid special functions defined with \eqref{Cfunc} as:
\begin{equation}
\,_{H}K_\mu^{(m)}(x, y | \xi) = \sum_{k = 0}^\infty \frac{\xi^k}{k!}\,_{H}C^{(2)}_{m k + \mu}(x, y) \qquad\qquad (\mu \in \mathbb{R}).
\end{equation}
In spite of their awkward aspect, these functions have quite simple properties and their study may shine further light on the properties of multivariable Bessel functions, which are playing a role of central importance in the theory of multi-photon processes induced by charged systems in intense laser fields \cite{ELotsteds08}. The same procedure can be applied to the derivation of sum rules involving products of three Bessel functions.

Also the evaluation of sums of the type \cite{GBevilacqua11}
\begin{equation}\label{eq19}
S_l^{(m)} (x, y) = \sum_{n = -\infty}^\infty n^m\,J_{n + l} (x)\,J_n (y), \qquad (m, l \in\mathbb{Z})
\end{equation}
required in some problems in plasma physics and quantum optics, can be performed quite straightforwardly using our symbolic method.  By noting that 
\begin{equation}\label{eq20}
S_l^{(m)} (x, y) = \left[ (-i \partial_{\vartheta})^m\,\sum_{n = -\infty}^\infty e^{i n \vartheta}J_{n + l}(x) J_n (y)\right]_{\vartheta = 0},  
\end{equation}
from the Graf's addition theorem one has
\begin{equation}\label{eq21}
\sum_{n = -\infty}^\infty e^{i n \vartheta} J_{n + l}(x)\, J_n (y) = \left(\hat{c}\,\frac{x - y\,e^{-i \vartheta}}2\right)^l\,\exp\left\{-\hat{c}\,\frac{x^2 + y^2 - 2\,x\,y\,\cos\vartheta}4\right\}\,\varphi_0,
\end{equation}
and by applying to this expression the so-called Hoppe formula for the derivative of composite functions \cite{WPJohnson02} 
\begin{equation}\label{eq23}
\frac{{\mathrm d}^m}{{\mathrm d} t^m}\,g\left(f(t)\right) = \sum_{k = 0}^m \frac{g^{(k)}\left( f(t) \right)}{k!}\, A_{m, k}\left(f(t)\right)
\end{equation}
with 
\begin{equation}
A_{m, k}\left(f(t)\right) = \sum_{j=0}^{k}\left(\begin{array}{c} {k}\\{j}\end{array}\right)\, \left[- f(t)\right]^{k-j} \frac{{\mathrm d}^m}{{\mathrm d} t^m}\left[f(t)\right]^j, \nonumber
\end{equation}
we finally obtain 
\begin{eqnarray}\label{eq24}
S_l^{(m)} (x, y) &=& (-i)^m\,\left(\frac{\hat{c}}2\right)^l\,e^{-\hat{c}\,(x^2 + y^2)/4}\,\left\{\partial_{\vartheta}^m \left[(x - y e^{-i \vartheta})^l\,
e^{\hat{c}\,(x\,y\,\cos \vartheta)/2}\right]\right\}_{\vartheta = 0}\,\varphi_0 \nonumber \\ 
&=& \sum_{j = 0}^m \binom{m}{j} \sum_{k = 0}^l \binom{l}{k}\,(k - l)^j \sum_{p = 0}^{m - j} \frac1{p!}\,R^{(l, m)}_{k, p} (x, y) 
\end{eqnarray}
where
\begin{equation}
R^{(l, m)}_{k, p} (x, y) = \sum_{q = 0}^p \binom{p}{q}\,\left(- \frac12\right)^q \sum_{r = 0}^q \binom{q}{r}\,
(2 r - q)^{m - j}\,\frac{x^{k + p}\,(- y)^{l - k + p}}{(x - y)^{l + p}}\,J_{l + p} (x - y).
\end{equation}
In a similar manner, by using Eq. \eqref{Besumrul} specialized to the case of integer $\nu$, we can write ($m, l \in\mathbb{Z}$)
\begin{eqnarray}\label{eq26}
E_l^{(m)} (x) &=& \sum_{n = 0}^\infty \frac{n^m}{n!}\,J_{n+l} (x) = \left[(t\,\partial_t)^m \sum_{n = 0}^\infty \frac{t^n}{n!}\,J_{n + l} (x) \right]_{t = 1} \nonumber \\ 
&=& \left[(t\,\partial_t)^m \left(\hat{c}\,\frac{x}2\right)^l\,\exp\left(- \hat{c}\,\frac{x^2 - 2\,x\,t}4\right)\,\varphi_0 \right]_{t = 1} 
\end{eqnarray}
and, by recalling that \cite{LCarlitz74, GDattoli12, KAPenson04}
\begin{equation}\label{eq27}
(t \partial_t)^m = \sum_{k = 1}^m S_2 (m, k)\,t^k\,\partial_t^k
\end{equation}
with $S_2 (k, m)$ Stirling numbers of the second kind, we find (see also \cite{GDattoli04-2})
\begin{eqnarray}\label{eq28}
E_l^{(m)} (x) &=& \sum_{k = 1}^m S_2 (m, k)\,\left(\hat{c}\,\frac{x}2\right)^{l + k}\,e^{- \hat{c}\,(x^{2} - 2\,x\,t)}\,\varphi_0 \nonumber \\
 &=&  \sum_{k = 1}^m S_2 (m, k)\,\left(\frac{x}2\right)^{l + k}\,C_{l + k} \left(\frac{x^2 - 2\,x\,t}4\right)
\end{eqnarray}
where $C_{l + k} $ is the Tricomi-Bessel function defined in Eq. \eqref{BesTri}. 

\noindent
All the new results obtained above have been verified numerically.

In this paper we have shown that using a symbolic procedure of umbral type, significant progress can be achieved in the theory of sum rules involving Bessel functions. Such a point of view has opened a new scenario 
offering a wealth of analytical possibilities amenable for further generalizations of practical interest in various fields involving the theory of special functions. We have indeed mentioned laser assisted 
processes in strong laser fields \cite{ELotsteds08}, but we should also include synchrotron radiation problems regarding the emission of electron radiation in nonconventional undulator magnets 
\cite{Iracane91, PChaix93, GDattoli93}. Within this last framework a significant role has been played by the so called multi-index Bessel functions \cite{Dattoli94}, whose theory can be greatly simplified by the umbral formalism 
discussed here.

%-----------------------------------------------------------

\section{Acknowledgements}

K. G. thanks Funda\c{c}\~{a}o de Amparo \'{a} Pesquisa do Estado de S\~{a}o Paulo (FAPESP, Brazil) under Program No.~2010/15698-5. K.~G. also thanks ENEA, Frascati, Roma, for financial support and kind hospitality. K. A. P. 
acknowledges support from Agence Nationale de la Recherche (Paris, France) under Program PHYSCOMB No.~ANR-08-BLAN-243-2. 

\appendix{}
\section{Proof of Eq. \eqref{Besumrul2}}
In this Appendix we prove the identity \eqref{Besumrul2}, used in sec. \ref{s:symbolic} to prove the Graf's addition theorem.
By starting from the familiar derivative formula \cite{GEAndrews01}
\begin{equation}
%\label{ }
\left(x^{- 1}\,\partial_x\right)^n\,\left[x^{\nu}\,J_\nu (x)\right] =  x^{\nu - n}\,J_{\nu - n} (x) \qquad \qquad (\nu\in\mathbb{C}, n = 0, 1, 2,\ldots),
\end{equation}
we obtain
\begin{equation}
%\label{ }
\exp\left\{\tau\,x^{- 1}\,\partial_x\right\}\,\left[x^{\nu}\,J_\nu (x)\right] = x^\nu\,\sum_{k = 0}^\infty \frac{(\tau/x)^k}{k!}\,J_{\nu - k} (x)
\end{equation}
and, by using the operational identity \cite{GDattoli97}
\begin{equation}
%\label{ }
\exp\left\{\tau\,x^{- 1}\,\partial_x\right\}\,f (x) = f (\sqrt{x^2 + 2\,\tau}), 
\end{equation}
we get
\begin{equation}
%\label{ }
\sum_{k = 0}^\infty \frac{(\tau/x)^k}{k!}\,J_{\nu - k} (x) = \left(\frac{\sqrt{x^2 + 2\,\tau}}{x}\right)^\nu\,J_\nu (\sqrt{x^2 + 2\,\tau})
\end{equation}
from which, after setting $\tau = - x\,t$, we recover \eqref{Besumrul2}.

%-----------------------------------------------------------

\end{document}